\documentclass[pra,twocolumn,nofootinbib]{revtex4-2}

\usepackage{dcolumn}% Align table columns on decimal point
\usepackage{bm}% bold math
\usepackage{mathtools}
\usepackage{siunitx}
\usepackage{graphicx}% Include figure files
\usepackage{amssymb,amsmath}
\usepackage[version=4]{mhchem}

\usepackage{booktabs}

\begin{document}
\newcommand{\dd}{\text{d}}
%\DeclareSIUnit{\molar}{M}

\newcommand{\kB}{k_{\text{B}}}
\newcommand{\kT}{\kB T}

\newcommand{\ceq}{c^\text{eq}}
\newcommand{\ctot}{c^\text{tot}}

\newcommand{\Js}{J^\text{s}}
\newcommand{\cs}{c^\text{s}}

\newcommand{\RQ}{{\cal{Q}}}
\newcommand{\KbyQ}{{K/\RQ}}

\newcommand{\kon}{k_{\text{on}}}
\newcommand{\koff}{k_{\text{off}}}
\newcommand{\KGTP}{K_{\text{GTP}}}
\newcommand{\QGTP}{\RQ_{\text{GTP}}}

\newcommand{\kC}{k}
\newcommand{\lC}{l}
\newcommand{\kD}{k}
\newcommand{\lD}{l}

\newcommand{\Ecoli}{{\it E.~coli}}

\title{Why life is hot}
\author{Tanja Schilling}
 \email{tanja.schilling@physik.uni-freiburg.de}

 \affiliation{Institute of Physics, University of Freiburg, Hermann-Herder-Stra\ss e 3, D-79104 Freiburg, Germany.}

\author{Patrick B.~Warren}%
\affiliation{%
 The Hartree Centre, STFC Daresbury Laboratory, Warrington, WA4 4AD, United Kingdom
}%

\author{Wilson Poon}%
\affiliation{%
 School of Physics and Astronomy, The University of Edinburgh, Peter Guthrie Tait Road, Edinburgh EH9 3FD, United Kingdom.
}%

\date{\today}% It is always \today, today,
%TC:ignore
\begin{abstract}
 The process of evolution by natural selection leads to phenotypes of increasing fitness. For cellular chemical reaction networks, this means optimising a variety of fitness functions such as robustness, precision, or sensitivity to external stimuli. We argue that these diverse goals can be achieved by a versatile, generic mechanism: coupling chemical reaction networks to reservoirs that are strongly out of equilibrium. Using theory and numerics we show that this mechanism of optimisation comes at the price of significant heat dissipation. We compute the heat flux caused by kinetic proofreading in {\it Escherichia coli} and show that it constitutes a significant fraction of the total heat flux experimentally measured in this model organism. 
We then demonstrate that the degree of optimality achievable saturates, and that Nature appears to operate near saturation despite high energetic costs. We argue that `life is hot' largely because of the need for a versatile mechanism to optimise a variety of fitness functions. 
\end{abstract}
%TC:endignore
\maketitle

%\tableofcontents

\section{Introduction}

Two features of life are typically taken as self evident: first that it is a non-equilibrium phenomenon, and second that it ubiquitously generates heat.
% \footnote{Note that a few bacteria apparently absorb heat when they grow~\cite{Liu2001}.}  
That `life is non-equilibrium' is frequently discussed, especially in the physics literature~\cite{Fang2019}, because non-equilibrium statistical mechanics continues to pose formidable challenges.
In contrast, that life dissipates heat is considered so banal that it seldom draws comments except in textbooks.

Interestingly, the most common textbook explanation of why life must dissipate heat, e.g., as given in successive editions of a well-known university text~\cite{Alberts2022}, is incorrect. Alberts et al.~explain that the cellular metabolism generates `order', which lowers entropy. (Schr\"odinger coined the term `negentropy' for this effect~\cite{Schrodinger2012}.) To satisfy the Second Law of Thermodynamics, enough heat must be exported from the cell to raise the entropy of the environment to give a net positive entropy change. However, numerical estimates of the total entropy reduction due to various ordering processes within a cell (polymerisation, compartmentalisation, etc.) show that this argument underpredicts the heat production by around two orders of magnitude~\cite{Davies2013}. Cells reject much more heat than is needed to compensate for their 'negentropy'. Why, then, does life ubiquitously produce heat?

Before we give an answer to this question, we need to set some preliminaries. First, metabolic rates are almost universal across a large range of organisms from all domains. They stay within a narrow band of 1--$\SI{100}{\watt\per\kilo\gram}$ for organisms spanning 20 orders of magnitude of body mass~\cite{Gavrilov2008}. Not all of the energy that an organism takes up from its environment is converted into heat -- clearly, living organisms need to perform chemical and mechanical work to survive -- but the heat flux amounts to 60\% to 90\% of the energy balance \cite{ballesteros2018thermodynamic,lamprecht2003calorimetry,zotin1990thermodynamic}. From a physicist's perspective this observation contains two aspects that require an explanation: the quasi-universal behaviour across all domains and the apparent `waste' of energy as heat. 

Second, note that only a small fraction of all organisms are endotherms (i.e.~regulate their own body temperatures). Most organisms dissipate heat without increasing their temperature. The dissipated heat is taken up by the surrounding medium which acts as a heat bath.\footnote{The heat conductivity $k$ within cells is similar to that of water, $k \approx 0.6 \,\text{W}\text{m}^{-1} \text{K}^{-1}$ \cite{kyoo2013thermal,inomata2026transparent}. 
Consider a sphere of diameter $2R$ embedded in a medium. If the sphere emits heat at a power density $q$ (i.e.~power per unit volume) and the thermal conductivity in the sphere is the same as in the surrounding medium, then $\Delta T \simeq {R^2q}/{k}$. For a source which emits $1\text{pW}$ in water, e.g.~a single \Ecoli\ cell in the exponential phase, the change in temperature at a distance of $1\,\mu\mathrm{m}$ is only about $\Delta T = 10^{-7}\,\text{K}$.   
For a concentrated cluster of cells (or a larger organism) with a heat flux of $1\,\text{pW}$ per cell, one would have to go to clusters comprising $10^9$ to $10^{12}$ cells (i.e.~millimeters in diameter in the case of \Ecoli) to reach $\Delta T >1\text{K}$ in the centre of the cluster. Moreover, the power density of most organisms is a factor $10^{-3}$ to $10^{-2}$ smaller than the one of \Ecoli. Thus they need to have a diameter of at least a few centimeters to reach $\Delta T > 1 \text{K}$.} Regulation of the body temperature is vital for some animals, but it cannot be the general explanation of heat dissipation in all life-forms.

We will argue that, instead, the cause for heat dissipation lies in the evolutionary pressure to optimise chemical reaction networks for a variety of functions.
Sometimes organisms need to be precise (e.g.~when translating mRNA into proteins); sometimes they need to produce superstructures with tightly-defined properties (e.g.~long actin filaments with a narrow length distribution); and sometimes they need to respond at speed to external stimuli (e.g.~heat shock). We show that adding phosphorylation-dephosphorylation cycles to chemical reaction networks is a versatile way to optimise them for a variety of fitness functions, but necessarily dissipates a significant amount of heat.

The article is structured as follows: First we discuss the role of cycles in the optimisation of chemical reaction networks. Then we compute the heat flux in two well-known biochemical processes. We then compare our prediction of the heat dissipated by kinetic proofreading in {\it Escherichia coli} to results of calorimetry experiments and conclude that proofreading causes a significant fraction of the total heat flux. Further, we show that the degree of optimality saturates, and that Nature appears to operate near saturation. Finally, we suggest an explanation for the observed quasi-universality of metabolic rates.

\section{The role of phosphorylation-dephosphorylation cycles in optimisation}

\begin{figure}
 \centering
    \includegraphics[width=0.9\linewidth]{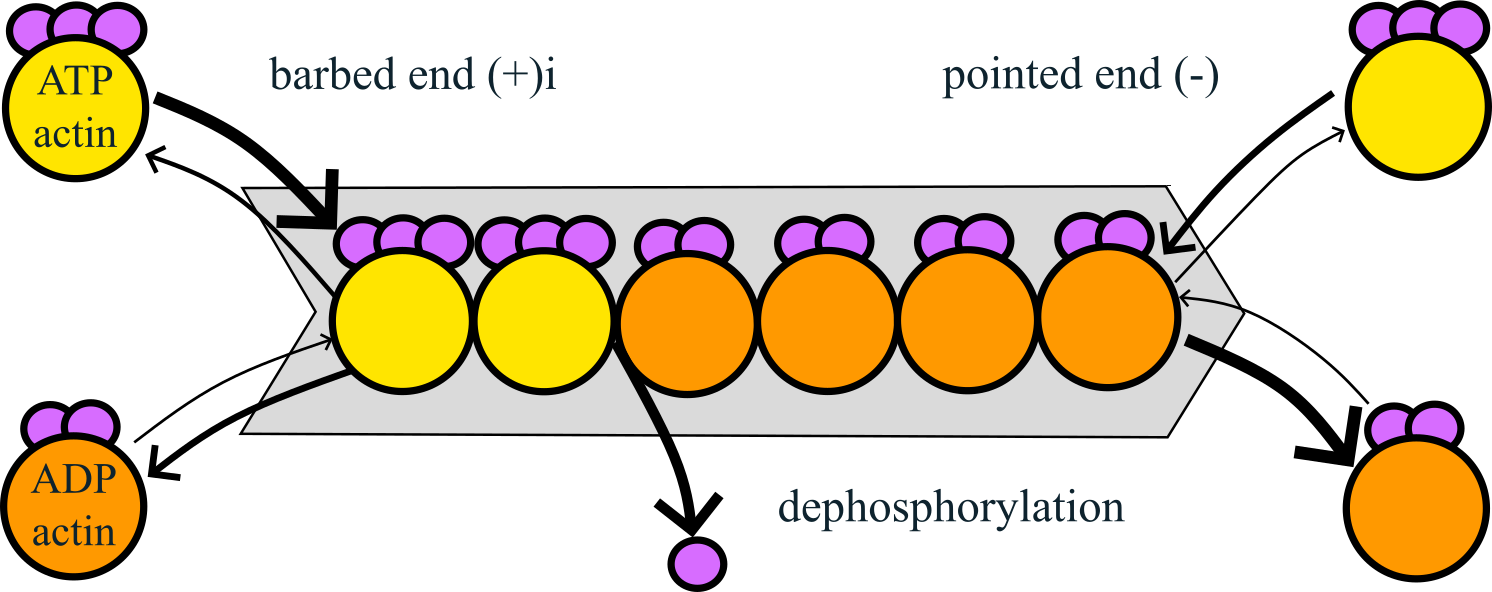}   
  \caption{\label{fig:ActinSketch} Sketch of actin polymerization. Yellow spheres denote ATP-actin, orange spheres denote ADP-actin, the individual purple sphere denotes
inorganic phosphate. The thickness of the arrows indicates the magnitude of the reaction rate.}
\end{figure}

We begin our discussion with an example of a biochemical optimisation problem that is solved by means of a driven reaction cycle. To enable the proper functioning of the cytoskeleton, actin filaments need to have a well-defined length \cite{pollard2000molecular}. If actin polymerisation proceeded under thermodynamic equilibrium conditions, there would be a large dispersity in filament lengths. In order to reduce the length dispersity, cells couple the polymerisation process to the dephosphorylation of ATP as sketched in Fig.~\ref{fig:ActinSketch} \cite{erlenkamper2013treadmilling}.

There are two types of `monomeric' actin, ATP bound (yellow spheres) and ADP bound (orange spheres). The ends of the filament have different structures (indicated by plus and minus signs). The rates of attachment and detachment differ between the ends as well as between the types of monomers (in the sketch the magnitude of the rate is indicated by the thickness of the arrow). If the dephosphorylation step were missing, none of these details would matter. The length distribution would  be a wide Poissonian. The dephosphorylation step, which changes the properties of the monomer units while they are part of the polymer, is strongly out of equilibrium and therefore affects the nature of the length distribution. In the so-called `treadmilling regime' the distribution becomes a convolution of a narrow Poissonian with a narrow Gaussian \cite{erlenkamper2013treadmilling,Xin2009,carlier1997control,neuhaus1983treadmilling,hadjivasiliou2023selection}. Thus the phosphorylation-dephosphorylation cycle reduces the dispersity in filament lengths.

Cycles of this form appear ubiquitously in biochemistry. Further examples are:
\begin{itemize}
\item In protein synthesis the error is reduced with respect to its equilibrium value by means of the Hopfield kinetic proofreading cycle \cite{Hopfield1974,ninio1975kinetic,Boeger2022}. Here, the quantity which is optimised is the ratio between two currents. (We discuss the details in sec.~\ref{sec:Results}~\ref{sec:kineticProofreading}.)
\item When heat shock proteins bind to the proteins that they refold, they show `ultra-affinity'. In this case a dephosphorylation cycle is used to enhance a binding affinity over its equilibrium value \cite{nguyen2017thermodynamic,de2014hsp70}.
\item In chemotaxis the sensitivity is optimised by a reaction cycle that is equivalent in structure to the Hopfield proofreading cycle \cite{hartich2015nonequilibrium}.
\end{itemize}

To see why such cycles are useful, we consider the mass action rate law for a network of chemical reactions. We denote by $c_i(t)$ the concentration of chemical species $i$ and by $k_{ij}$ and the reaction rate coefficient between species $i$ and $j$. The concentrations evolve as 
\begin{equation}
\label{eq:Master}
\frac{\dd c_i(t)}{\dd t} = \sum_j \left(k_{ij}c_j(t)-k_{ji} c_i(t) \right) = \sum_i J_{ij}\quad ,
\end{equation}
where we have defined the currents $J_{ij}(t)\equiv k_{ij}c_j(t)-k_{ji}c_i(t)$.
We are, in particular, interested in the steady state solutions of eq.~\eqref{eq:Master}, $c_i(t) = \text{const} \equiv \cs_i$. 
%Later we will analyze reaction networks that contain enzymes the concentrations of which vary slowly compared to the speed of the other reactions. 

\begin{figure}
 \centering
    \includegraphics[width=\linewidth]{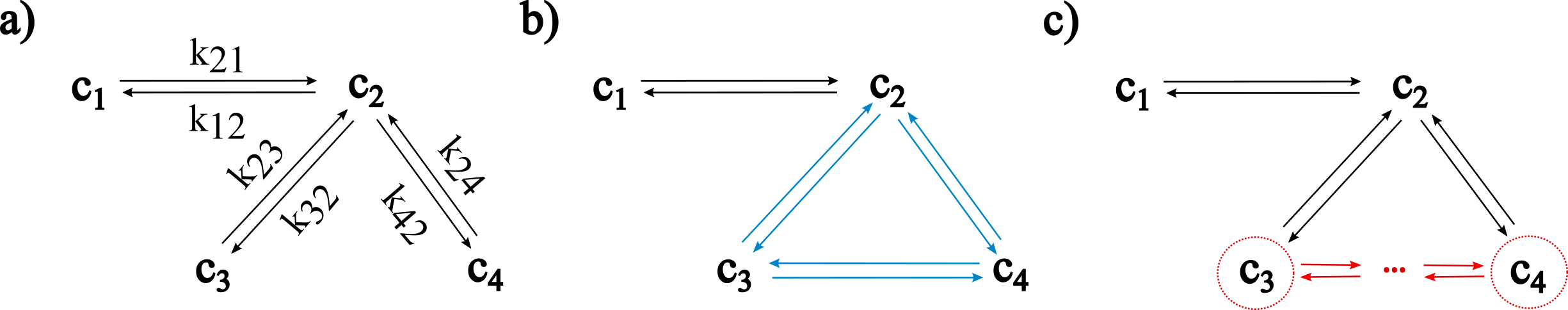}   
  \caption{\label{fig:NetworksSketch} Illustration of mass action rate laws in terms of graphs. Vertices represent concentrations, edges represent rate coefficients. a) A tree-like graph, i.e.~a graph that does not contain cycles. b) The nodes $c_2$, $c_3$ and $c_4$ form a cycle (blue arrows). We removed the labels for the rate coefficients to avoid cluttering the graph. c) The concentrations at the nodes $c_3$ and $c_4$ are fixed by means of particle reservoirs (indicated by the red dashed circles). This creates a cycle indirectly (red arrows), because the reservoirs need to be maintained by means of another chemical reaction network (indicated by the ellipsis).}
\end{figure}

Chemical reaction networks can be represented by weighted graphs (see fig.\ref{fig:NetworksSketch}). The concentrations $c_i$ are the vertices of such a graph and the rate coefficients are the edges. In graph theory, the term "cycle" refers to a set of edges that connect a vertex to itself.
Fig.~\ref{fig:NetworksSketch} a) shows a "tree-like" graph, i.e.~a graph that does not contain any cycles. In fig.~\ref{fig:NetworksSketch} b) the vertices $c_2$, $c_3$ and $c_4$ form a cycle (blue arrows). In fig.~\ref{fig:NetworksSketch} b) the concentrations at the vertices $c_3$ and $c_4$ are fixed by means of particle reservoirs (indicated by the red dashed circles). This creates a cycle indirectly (red arrows), because the reservoirs need to be maintained by means of another chemical reaction network (indicated by the red ellipsis).

Consider now how we may optimise the steady state of a network with respect to some fitness function. To give a few examples, this function could be the dispersity of the values $\cs_i$ (as in the case of the actin filaments), the speed with which a certain particle species is produced, the average time two species remain in a complex, or the response of a concentration to a small change in a current, $\partial\cs_i/ \partial\Js_{jk}$. 

If the network is tree-like, the steady-state concentration ratios $\cs_i/\cs_j$ are given by the detailed balance condition $\Js_{ij} = k_{ij}\cs_j-k_{ji}\cs_i =0 \; \text{for all}\; i,j$ \cite{horn1972general}. Thus the absence of cycles significantly limits the set of functions that can be optimised. To change the concentration ratios in a tree-like network, one must either alter rate coefficients $k_{ij}$ or add species and therefore reactions to the network. Both mechanisms occur in living organisms, e.g., via point mutations to extant proteins or the emergence of (say) a novel inhibitor.\footnote{We use the term 'novel' here to mean a species not already present in the network. It could of course be an existing inhibitor or compound that is co-opted to a new role.} However, either mechanism requires a one-off, specific chemical innovation as each new challenge arises; neither is a suitable `design principle' for a generic solution.

To design a generic solution for optimising networks for multiple fitness functions without changing the chemistry of the constituents, we need to include cycles. 
In a network with cycles, the steady-state currents $\Js_{i,j}$ can be different from zero, thus
in particular, the sensitivities ${\partial \ln \cs_i}/{\partial \ln \Js_{jk}}$ can be optimised.
The extreme case of optimisation by means of a cycle concerns the so-called `futile' cycles in which a chemical species is disassembled into its building blocks only to be reassembled again without any obvious purpose. However, if a `futile' cycle is attached to a reaction network, the steady state concentrations are shifted away from their equilibrium values and thus the sensitivity of the network to the boundary conditions is altered. As has been previously pointed out, the cycle is not at all futile if it is placed at the right node to tune the network's response to variations in currents or concentrations \cite{qian2006metabolic, qian2007phosphorylation,beard2008chemical}.

In living organisms, perhaps the most versatile cycles are those involving the dephosphorylation of `energy currency' metabolites such as ATP, which are chemiostatted\footnote{In the language of stochastic thermodynamics, fixing concentrations by means of reservoirs is called 'chemiostatting'   \cite{seifert2011stochastic}.} at concentrations very far from their equilibrium values~\cite{mrnjavac2025gtp}.  According to biology textbooks, ATP is used for the following purposes: to enable reactions that are energetically unfavourable, to transport substances across membranes, and to do chemical or mechanical work \cite{Alberts2022}. We argue that ATP has a fourth and at least equally important task: to optimise the emergent properties of chemical reaction networks by driving them strongly out of equilibrium.

A large variety of fitness functions can be tuned by means of NTP-ases (i.e.~enzymes that catalyse the dephosphorylation of nucleoside-triphosphates such as ATP) which couple phosphorylation-dephosphorylation cycles into chemical reactions. However, this advantage comes at a cost that is manifested as a ubiquitous characteristic of life: a significant amount of heat is produced.

\section{Results}
\label{sec:Results}
\subsection{Dissipation in chemiostatted biochemical networks}
\begin{figure}
 \centering
    \includegraphics[width=0.9\linewidth]{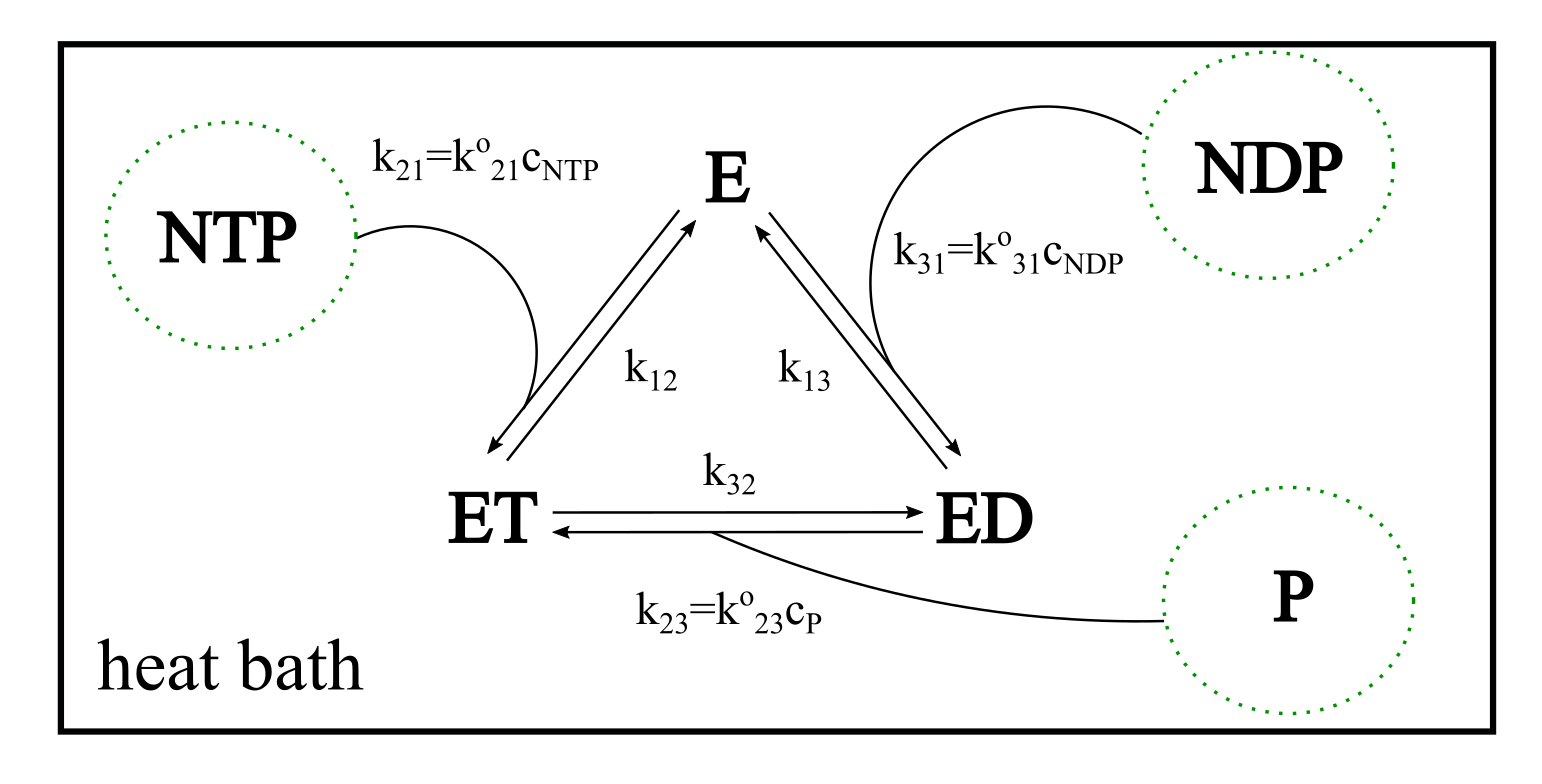}   
  \caption{\label{fig:network}Reaction network coupled to a heat bath, which fixes the temperature $T$, and to three particle reservoirs (indicated by dashed circles), which control the concentrations of the species $\text{NTP}$, $\text{NDP}$ and $\text{P}$.}
\end{figure}

To estimate the amount of heat that is produced, consider a chemical reaction network which is attached to three particle reservoirs and embedded in a heat bath at a temperature $T$, Fig.~\ref{fig:network}.\footnote{In real biochemistry, such a network will be part of a larger metabolic pathway. We analyze this as a self-contained structure only as a minimal model to estimate heat production in driven reactions. The letters NTP do not refer to a specific species of molecule, but are a place holder for ATP, GTP, other nucleoside triphosphates or complexes formed with any of these molecules. We will specify details later in the examples.} 
The system cycles through the states E, ET and ED transporting molecules from the NTP-reservoir to the NDP- and P-reservoirs, i.e.~the species E acts as a catalyst for the reaction 
\begin{equation} 
\ce{NTP}\rightleftharpoons\ce{NDP + P}\,.
\label{eq:APD}
\end{equation} 
Species $E$ takes part in three reactions 
\begin{equation*}
\ce{A + E} \xrightleftharpoons[k_{12}]{k^0_{21}} \ce{ET} \; , \quad 
\ce{ET} \xrightleftharpoons[k^0_{23}]{k_{32}} \ce{ED + P}\; , \quad 
\ce{ED} \xrightleftharpoons[k^0_{31}]{k_{13}} \ce{E + D}\; ,
\end{equation*}
where $k_{ij}$ are first order rate coefficients and $k_{ij}^0$ are second order rate coefficients, i.e.~the corresponding first order coefficients are given by products with concentrations such as $k_{21}^0c_\text{NTP} =: k_{21}$.
For the cyclic process to be possible in equilibrium, the reaction rate coefficients need to fulfill the condition 
\begin{equation}
\frac{k^0_{21}k_{32}k_{13}}{k_{12}k^0_{23}k^0_{31}} 
= \frac{\ceq_\text{NDP}\ceq_\text{P}}{\ceq_\text{NTP}} \, ,
\end{equation}
where $\ceq_i$ is the concentration of species $i$ in equilibrium.\footnote{To see this, consider the special case in which the concentrations in the reservoirs are fixed to the equilibrium values of Reaction~(\ref{eq:APD}). Then the product of the ratios of forward to backward rates $(k^0_{21}\,\ceq_\text{NTP}/k_{12})\cdot(k_{32}/(k^0_{23}\,\ceq_\text{P}))\cdot(k_{13}/(k^0_{31}\,\ceq_\text{NDP}))$ has to equal $1$.}

If we set the concentrations in the NTP, NDP and P reservoirs to values other than the equilibrium concentrations, the system runs in a non-equilibrium steady state (NESS).
The change of free energy $\Delta G$ per mole of reaction in the NESS is determined solely by the properties of NTP, NDP and P
\begin{equation}
   \frac{\Delta G}{k_\text{B}T} = \ln\left(\frac{\RQ}{K}\right)\; ,
   \label{eq:DeltaG}
\end{equation} 
where $k_\text{B}$ is Boltzmann's constant, $\RQ=c_\text{NDP}c_\text{P}/c_\text{NTP}$ is the reaction quotient, $c_\text{NDP}$, $c_\text{P}$ and $c_\text{NTP}$ denote the reservoir concentrations, and $K = {\ceq_\text{NDP}\ceq_\text{P}}/{\ceq_\text{NTP}}$ is the equilibrium constant for Reaction~(\ref{eq:APD}).
(Note that we use two similar looking letters: $\RQ$ for the reaction quotient and $Q$ for heat.)

In the NESS entropy is produced at a rate \cite{beard2008chemical,schnakenberg1976network}
\begin{equation}
\frac{\dd S^\text{sys}(t)}{\dd t} =  \frac{\kB}{2} \sum_{i,j\ne i}\left[\left(k_{ji}\,\cs_i - k_{ij}\,\cs_j \right) \ln \left( \frac{k_{ij}\cs_j}{k_{ji}\cs_i}\right)\right] ,
\label{eq:entropy}
\end{equation} 
where $i,j\in \{1, 2, 3 \}$ label the states $\{\ce{E}, \ce{ET}, \ce{ED} \}$.\footnote{We have used the notation ${\dd S^\text{sys}}\!/{\dd t}$ for the change in entropy that is not due to heat exchange, while the biochemical literature often follows Prigogine and uses the notation ${\dd_\text{i} S(t)}/{\dd t}$ for this quantity \cite{Prigogine2014}.} % The other part of the change in entropy, which is due to heat exchange, is called "entropy flow" in the literature on stochastic thermodynamics \cite{boeger2022kinetic, beard2008chemical}.}

Eq.~\eqref{eq:entropy} can be simplified. First, we note that both factors in the sum are antisymmetric under exchange of the indices, i.e.~$(k_{ji}\,\cs_i - k_{ij}\,\cs_j) = - (k_{ij}\,\cs_j - k_{ji}\,\cs_i)$ and $\ln \left( \frac{k_{ij}\cs_j}{k_{ji}\cs_i}\right) = - \ln \left( \frac{k_{ji}\cs_i}{k_{ij}\cs_j}\right)$. Thus we can replace the sum over $i$ and $j$ by twice the sum over the pairs $(1,2),(2,3)$ and $(3,1)$. Second, in the steady-state the net current in the cycle, i.e.~the difference between the anti-clockwise and the clockwise currents, equals $\Js \equiv \left(k_{21}\,\cs_1 - k_{12}\,\cs_2 \right)= \left(k_{32}\,\cs_2 - k_{23}\,\cs_3 \right) = \left(k_{13}\,\cs_3 - k_{31}\,\cs_1 \right)$. Third,

\begin{multline*}
\frac{\Delta G}{\kB T} = \ln\left(\frac{\mathcal{Q}}{K}\right) = \ln\left(\frac{c_\text{NDP}\,c_\text{P}}{c_\text{NTP}}\frac{\ceq_\text{NTP}}{\ceq_\text{NDP}\,\ceq_\text{P}}\right)\\ = \ln\left(\frac{c_\text{NDP}\,c_\text{P}}{c_\text{NTP}} \frac{k_{12}k^0_{23}k^0_{31}}{k^0_{21}k_{32}k_{13}} \right) = \ln\left(\frac{k_{12}\,k_{23}\,k_{31}}{k_{21}\,k_{32}\,k_{13}}\right)\,.
\end{multline*}

With this we obtain 
\[
\frac{\dd S^\text{sys}(t)}{\dd t} = \frac{1}{T}\,\Js\Delta G\; ,
\]

As we are studying a cyclic process of a system that is embedded in a heat bath, the heat flux from the system to the bath is given by \cite{Callen1985}\footnote{Note that the particle reservoirs are assumed to be perfect, i.e.~to keep strictly constant concentrations. If the particle number in the reservoirs were allowed to fluctuate, some of the entropy would enter the particle reservoirs and would not be detected in a calorimetry experiment. For a discussion of this issue see reference~\cite{seifert2011stochastic}, and for an exhaustive analysis of the stochastic thermodynamics of chemical reaction networks with particle reservoirs see reference~\cite{rao2018conservation}. As we are interested in estimating orders of magnitude, we focus on the idealized case of perfect chemiostatting.}
\begin{equation}
  \label{eq:heatfinal}
\frac{\dd Q}{\dd t} = T\frac{\dd S^\text{sys}(t)}{\dd t} = \Js \,\Delta G \, .
\end{equation}

Depending on the specific `energy currency', the temperature, pH and other conditions, $K\simeq 10^5$--$10^6\,\mathrm{M}$ \cite{rosing1972value,kotting2004time,alberty2005thermodynamics}. In cells, the concentration of energy sources containing ATP or GTP as well as the concentration of inorganic phosphate is in the millimolar range, while the concentrations of hydrolysed forms (ADP etc.) is typically in the micromolar range \cite{milo2010bionumbers}. Hence, in a biological organism, the reaction quotient $\RQ=c_\text{NDP}c_\text{P}/c_\text{NTP}$ is of the order $10^{-5}$--$10^{-6}\,\mathrm{M}$, which corresponds to $K/\RQ\sim10^{9}$--$10^{12}$, or equivalently
$\Delta G \simeq 21$--$28\,\kT\simeq50$--$\SI{70}{\kilo\joule\per\mol}$.
%\[\Delta G \simeq \qtyrange{50}{70}{\kilo\joule\per\mol}\, .\] 
Values of $K/\mathcal{Q}$ in this range are very large compared to values of this quotient for most cellular reactions. So, for example, for 24 reactions in the central carbon metabolism of {\it E.~coli}, 15 have $10^0 < K/\mathcal{Q} < 10^1$, and all but one have $ K/\mathcal{Q} \lesssim 10^4$ (see Appendix~\ref{app:sec:K/Q}). After providing some numerical estimates of heat production in biochemical reactions, we will propose an explanation of why natural selection has `tuned' $K/\RQ$ for ATP/GTP hydrolysis to such large values.

\subsection{1st Example: actin treadmilling}
\begin{figure}
 \centering
    \includegraphics[width=0.9\linewidth]{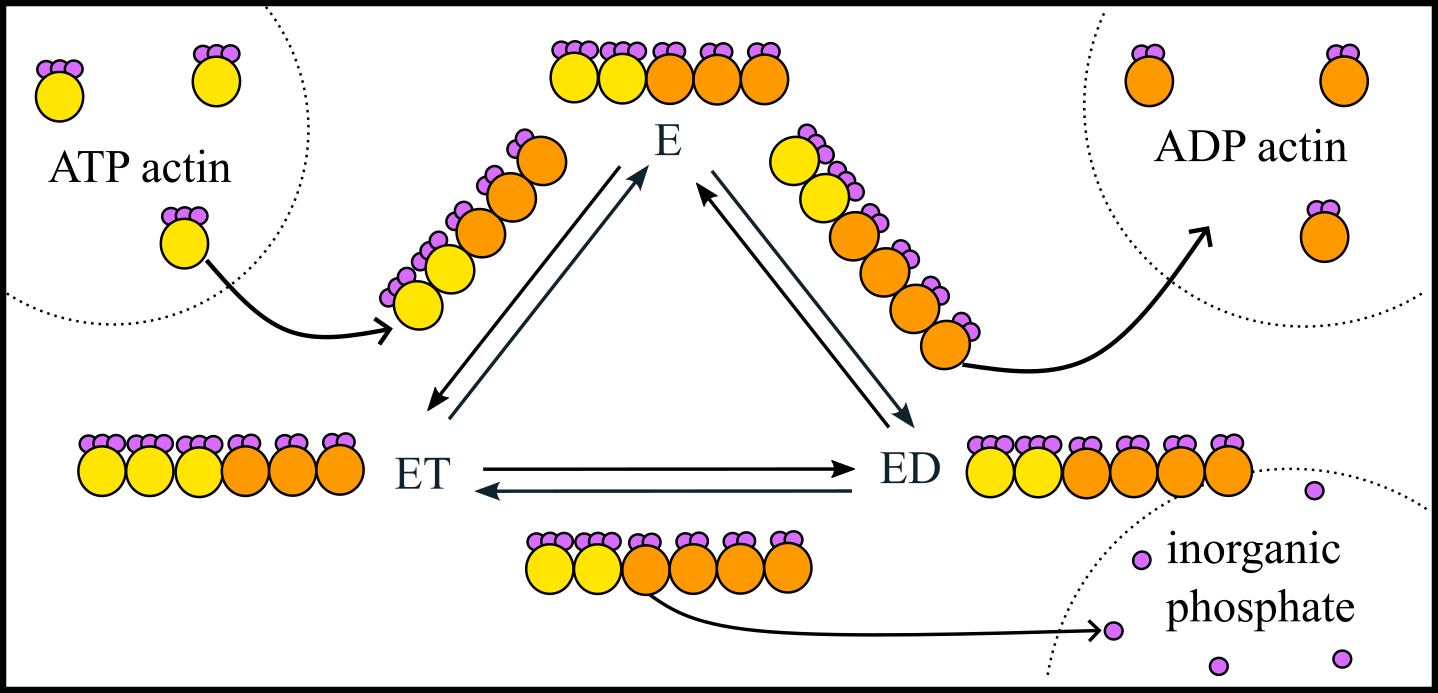}   
  \caption{\label{fig:Actin} Actin treadmilling. Yellow spheres with three purple spheres denote ATP-actin, orange spheres with two purple spheres denote ADP-actin, individual purple spheres denote inorganic phosphate.}
\end{figure}

To estimate the magnitude of heat dissipation, Eq.~\eqref{eq:heatfinal}, for actual biochemistry, we consider first actin treadmilling (Fig.~\ref{fig:Actin})  \cite{neuhaus1983treadmilling,hadjivasiliou2023selection, erlenkamper2013treadmilling,Xin2009,pantaloni2001mechanism}. The chain of bound F-actin subunits corresponds to species E. First, G-actin momomers are activated by binding to ATP (yellow spheres in Fig.~\ref{fig:Actin}). Activated monomers are then added to the `barbed' end of the filament ($\ce{E}\rightleftharpoons\ce{ET}$). While they are part of the filament they are hydrolyzed ($\ce{ET}\rightleftharpoons\ce{ED}$), and finally they drop off at the `pointed' end ($\ce{ED}\rightleftharpoons\ce{E}$). As the hydrolysis step is strongly irreversible, the system enters the  NESS known as `treadmilling' \cite{erlenkamper2013treadmilling}. 

Steady state fluxes in this process are on the order of $10$ subunits per second added to/removed from a filament \cite{carlier1997control,Xin2009,carlier1998control}. 
Hence we expect the heat production rate of a single actin filament to be of the order
\begin{equation*}
\left(\frac{\dd Q}{\dd t}\right)_\text{filament} \simeq \frac{10 \, \mathrm{ s}^{-1}\times60\,\mathrm{kJ\,mol}^{-1}}{6\times 10^{23}\,\mathrm{mol}^{-1}} \simeq 10^{-18}\,\mathrm{W}\,.%  \quad .
\end{equation*}
There are considerable spatiotemporal variations in actin filament concentration in cells. It is particularly high, $\sim 1\,\mathrm{mM}$, in the lamellipodia of migrating fibroblasts~\cite{pollard2000molecular,abraham1999actin}. 
Assuming that the filaments have a length of order $10^2$ subunits \cite{schaub2007analysis}, we expect to see a heat production of 
\begin{equation}
\left(\frac{\dd Q}{\dd t}\right)_\text{lamellipodia} \simeq 5 \, \mathrm{W}\,{\mathrm{kg}^{-1}} \;,
\end{equation}
 which lies within the range of measured values for the metabolic rate in a diversity of organisms that we referred to in the introduction~\cite{Makarieva2006,Gavrilov2008}. More detailed experimental testing of our estimate could perhaps come from calorimetric measurements of fragments shed from fish karatocytes that are essentially pieces of `pure cytoskeleton' engaged in lamellopodia-like motion~\cite{mogilner2020}.

\subsection{2nd Example: kinetic proofreading}
\label{sec:kineticProofreading}
\begin{figure}
 \centering
    \includegraphics[width=\linewidth]{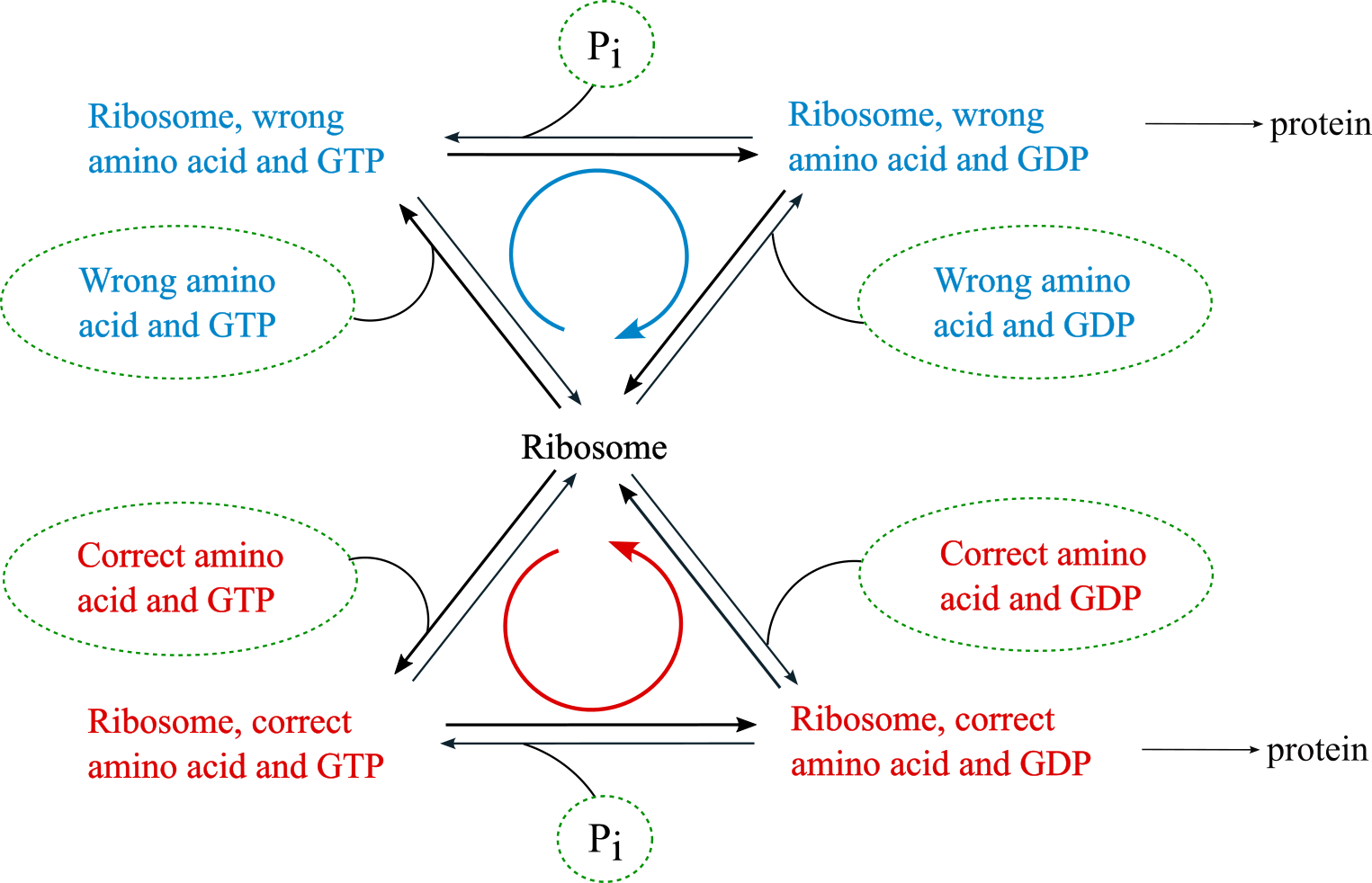}   
  \caption{\label{fig:proofreadingSketch} Sketch of the kinetic proofreading cycle \cite{Hopfield1974,ninio1975kinetic}. The reaction network consists of two cycles of the form shown in Fig.~\ref{fig:network}, one for the correct amino acid (red) and one for the wrong amino acid (blue). The fitness function that is optimised is the ratio of the 'wrong current' to the 'correct current' at the node of the ribosome.}
\end{figure}

An even more ubiquitous example of a chemiostatted cycle is the kinetic proofreading cycle (Fig.~\ref{fig:proofreadingSketch}), which regulates the error in translating mRNA to proteins in all organisms \cite{Hopfield1974,ninio1975kinetic}. To make numerical estimates, we consider proofreading in \Ecoli. In this case the species E from Fig.~\ref{fig:network} corresponds to the ribosome and the species NTP to the elongation factor complex between aminoacyl-tRNA and GTP. In the stationary (non-growing) phase, \Ecoli\ contains between $5000$ and $10000$ ribosomes  per cell (i.e.~about $10^{-20}\, \mathrm{mol}$). In the exponential growth phase there are up to 55\,000 ribosomes (i.e.~about $10^{-19}\,\mathrm{mol}$) \cite{dai2016reduction}.
The step $\ce{ET}\rightleftharpoons\ce{ED}$ consists of GTP activation and hydrolysis. Experimentally-determined rate coefficients for the combination of these processes vary between $k_{32}\simeq25\,\mathrm{s}^{-1}$ \cite{zaher2010hyperaccurate} and $54\,\mathrm{s}^{-1}$ \cite{rodnina2005recognition}. We know of no measured value of $k_{13}$, and so will follow the strategy of Zuckerman \cite{physicallens,physicallens2} and use the measured value of $k_{12}$ as an estimate, $k_{13} \simeq k_{12} \simeq 100\, \mathrm{s}^{-1}$ \cite{rodnina2005recognition}. These constants give us $k^\text{s} \simeq 20$--$35\,\mathrm{s}^{-1}$, which in turn yields a heat production rate per bacterium of 
\begin{equation}
\left(\frac{\dd Q}{\dd t}\right)_\text{kinpr}\simeq
\begin{cases}
0.2\,\mathrm{pW}&\text{(exponential phase)}\,, \\
0.02\,\mathrm{pW}&\text{(stationary phase)}\,.
\end{cases}\label{eq:ecoliq}
\end{equation}

As an \Ecoli\ bacterium has a mass of about \SI{1}{\pico\gram} \cite{kubitschek1986determination}, these values  correspond to 
%$\left(\frac{\dd Q}{\dd t}\right)_\text{kinpr}
$\left({\dd Q}/{\dd t}\right)_\text{kinpr}\simeq \SI{200}{\watt\per\kilo\gram}$ 
in the exponential phase and 
%$\left(\frac{\dd Q}{\dd t}\right)_\text{kinpr}
$\left({\dd Q}/{\dd t}\right)_\text{kinpr}\simeq \SI{20}{\watt\per\kilo\gram}$ 
in the stationary phase, which again lies in the range of measured values given in ref.~\cite{Makarieva2006,Gavrilov2008}.

The heat dissipation rate of actively growing \Ecoli\ has been measured directly by calorimetry. In the exponential phase, \Ecoli\ in various nutrient broths dissipate $\simeq 1$--4~pW per cell~\cite{Guisset2005}, so that kinetic proofreading, eq.~\eqref{eq:ecoliq}, is apparently responsible for a significant fraction of this dissipation. This seems reasonable, given that protein synthesis, of which kinetic proofreading is a key component, has been estimated to consume up to two-thirds of the energy budget of an active growing \Ecoli\ cell~\cite{Tempest1996}.

It is instructive to inquire into the origins of the remainder of the $\rm \gtrsim 1~pW$ heat dissipation in actively growing \Ecoli. To do so, we consider how much heat is produced by the process of chemiostatting, i.e.~of maintaining the ATP reservoir. For this we turn to two highly-curated models of the metabolism for \Ecoli\  developed for use within the framework of flux-balance analysis (FBA)~\cite{Palsson2010, palsson2015} (see Appendix section~\ref{app:sec:FBAAnalysis}). 

For fast-growing \Ecoli\ (doubling time of order 40--$50\,\mathrm{min}$) on a glucose minimal medium under aerobic conditions, we determined the total ATP maintenance demand is 60--$80\,\mathrm{mmol}$ per gram dry weight (gDW) of biomass, of which 6--10\% is for so-called non-growth-associated maintenance (NGAM).  Assuming all of this ATP is subsequently hydrolysed in `futile' cycles, this corresponds to a heat production rate of 0.3--$0.4\,\mathrm{pW}$ per bacterium.
% , which is comparable to the heat produced by kinetic proofreading in exponential growth in Eq.~\eqref{eq:ecoliq}.  
A similar conclusion is reached if the NGAM component, contributing 0.02--$0.04\,\mathrm{pW}$ per bacterium, is interpreted as predicting the heat production for maintaining the ATP reservoir concentrations in the stationary phase.\footnote{These estimates are admittedly rather crude, since we use the ATP maintenance demand as a proxy for other currency metabolites like GTP and we neglect for example the existence of overflow metabolism and other `sinks' for the ATP.  In principle, the calculation could be refined to include these effects. However, precision to this level is beyond the scope of the present work.}

We conclude that the heat dissipated by the NESS in the proofreading cycle and by refilling the reservoirs makes a significant or even dominant contribution to the overall heat production in both exponential growth and stationary phase.

\subsection{The steady-state current}
\label{sec:Current}
Finally, we address the question why Nature maintains the reservoir concentrations of nucleoside triphosphates so far from their equilibrium values, i.e.~$-\Delta G$ at such a high value. To do so, we return to the steady state current $\Js = k^\text{s}\ctot_\text{E}$ with $\ctot_\text{E} = \cs_{\text{E}} + \cs_{\text{ET}} + \cs_{\text{ED}}$ (see Appendix for details of the derivation). 
The rate coefficient $k^\text{s}$ is the difference between the product of the `anticlockwise rate coefficients', $k_{21}k_{32}k_{13}$, and the product of the `clockwise rate coefficients', $k_{12}k_{23}k_{31}$ 

\begin{equation}
k^\text{s} = \frac{k_{21}k_{32}k_{13}- k_{12}k_{23}k_{31}}{d}\; ,
\label{eq:steadystaterate}
\end{equation}
normalized by
\begin{equation*}
\begin{split}
&d = k_{21}k_{32} + k_{32}k_{13} + k_{13}k_{21} + k_{31}k_{23} + k_{23}k_{12}\\ 
&\hspace{3em}{} + k_{12}k_{31} + k_{13}k_{12}  +  k_{21}k_{23}  + k_{32}k_{31}\;,
\end{split}
\end{equation*}
where we have set 
$k_{21} = k^0_{21}c_\text{NTP}$, $k_{23} = k^0_{23}c_\text{P}$, and $k_{31} = k^0_{31}c_\text{NDP}$. 

\begin{figure}
 \centering
  % note on generating this figure
  % first create a python venv, then install copasi-basico and whatever else is needed, then do
  % /path/to/venv/python ./kpr_full_model_chemiostatted.py -t -o kpr_plot.pdf --dpi 300 -aaa
  % include a '-t' option switches to Times New Roman more or less
  % pdftoppm kpr_plot.pdf -png > kpr_plot.png
  \includegraphics[width=0.9\linewidth]{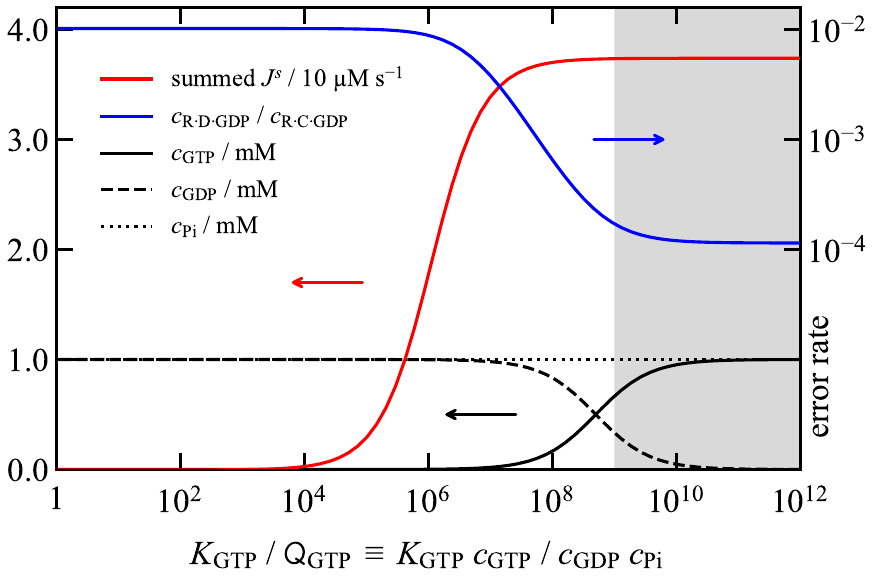}   
  \caption{ \label{fig:kpr}Kinetic proof reading model: summed cycle flux (red) and error rate (blue), expressed as a function of $\KbyQ$ for the GTP hydrolysis equilibrium. 
  %(Square brackets indicate concentrations.) 
  The error rates saturates at $\KbyQ \simeq 10^{10}$.  The shaded area indicates the range of $\KbyQ$ observed in living organisms \cite{milo2010bionumbers}.
  For the calculation, the GTP\,:\,GDP ratio is varied, constraining $c_{\text{GTP}} + c_{\text{GDP}}$ and $c_\text{Pi}$ to $1\,\text{mM}$.  Elongation factor complex concentrations are $0.1\,\text{mM}$ for both `C' (correct) and `D' (wrong) amino acid types.  A copy number of $5000$ ribosomes is assumed in a cell volume $\SI{2}{\micro\meter^3}$ (concentration $\simeq4.2\,\mu\text{M}$). Detachment rates of the complexes are enhanced by a factor $f=100$ for the `wrong' amino acid type.  For the saturated total cycle flux $\Js/[\ce{R}]_{\text{tot}}=k^\text{s}\simeq10\,\mathrm{s}^{-1}$, similar to the range quoted in the main text above \eqref{eq:ecoliq}. For more details of the model see Appendix section~\ref{app:sec:KPRModel}.}
\end{figure}

If the flux is predominantly anticlockwise (${k_{21}k_{32}k_{13} \gg k_{12}k_{23}k_{31}}$), re-phosphorylation is negligible ($k^0_{23}c_\text{P}\ll k^0_{21}c_\text{NTP}$) and re-attachment of the hydrolysed complex is also small ($k^0_{31}c_\text{NDP}\ll k^0_{21}c_\text{NTP}$), the steady state current approaches
\begin{equation}
\label{eq:Jlim}
\Js_\text{lim} = \frac{k_{32}k_{13}}{k_{32} + k_{13} + k^0_{23}c_\text{P}} \; \ctot_\text{E}\; ,
\end{equation}
i.e.~it saturates at a value that is independent of the reservoir concentrations $c_\text{NTP}$ and $c_\text{NDP}$. As a consequence, chemiostatting the reservoir concentrations even further out of equilibrium would not change the values of any fitness functions anymore. This saturation effect has been observed in the translation error in kinetic proof reading, which cannot be reduced beyond a certain limit per proof-reading cycle \cite{murugan2012speed,qian2006metabolic}, and in ultra-sensitivity \cite{hartich2015nonequilibrium}. Generally, when using a cycle of the type shown in Fig.~\ref{fig:network} in order to optimise a fitness function, there is a reaction quotient, beyond which nothing is gained anymore by converting even more chemical energy into heat. 

To explore this conclusion quantitatively, we implemented a chemiostatted version of a previously-described proofreading model~\cite{banerjee2017elucidating, physicallens}, which includes a cycle for the attachment of the correct amino acid to the growing polypeptide chain, as well as a prototypical cycle for the mistaken attachment of a `wrong' amino acid; in the latter the elongation factor complex detachment rates are enhanced.  We used rate coefficients and concentrations representative of the situation described above (see Appendix section~\ref{app:sec:KPRModel} for details).  

Figure~\ref{fig:kpr} shows the cycle flux summed over both cycles and the error rate (i.e.~the rate of attachment of `wrong' amino acid versus the correct amino acid) as a function of the disequilibrium measure $\KbyQ$ for $\ce{GTP}\rightleftharpoons\ce{GDP + Pi}$. The shaded area in Fig.~\ref{fig:kpr} indicates the range of $\KbyQ$ values observed in living organisms. The proofreading effect is clearly seen in the dramatic drop off in the error rate (blue line) for $\KbyQ\gtrsim10^7$. The summed cycle flux (red line), which is dominated by the attachment of correct amino acids, appears to saturate earlier than the proofreading effect. This can be traced to the delayed saturation of the `wrong' attachment cycle. Optimal proofreading requires $\KbyQ\gtrsim 10^{10}$. Interestingly, the shaded region begins at around this value. This simple model therefore explains the physiological requirement for such a large value of $\Delta G$ for the fueling currency metabolite (GTP), with the accompanying large heat production arising from its continual consumption.

Before we conclude, we briefly return to the question why the metabolic rates of a large variety of organisms are almost universal. We recall that the concentration of inorganic phosphate is usually in the micromolar range, thus the term $k^0_{23}c_\text{P}$ in $\Js_\text{lim}$ (eq.~\eqref{eq:Jlim}) is negligible and the rate coefficient $k^\text{s}$ approximately equals 
\begin{equation}
k^\text{s}_\text{lim}\simeq \frac{k_{32}k_{13}}{k_{32} + k_{13}}\; ,
\end{equation}
which is the harmonic mean of the rate coefficient of dephosphorylation and the rate coefficient by which the hydrolysed complex drops off the enzyme. These rate coefficients do vary between different nucleoside-triphosphatases, but not by many orders of magnitude. Hence it stands to reason that the observed values for the heat, eq.~\ref{eq:heatfinal}, stay in a limited range. 

\section{Summary and Conclusion}
We have presented three hypotheses:
\begin{enumerate}   
    \item To couple chemiostatted phosphorylation-dephosphorylation cycles into chemical reaction networks is a versatile mechanism to optimise fitness functions.
    \item Much of the heat produced by biological organisms is a necessary byproduct of this mechanism. 
    \item Metabolic rates are quasi-universal across all domains, because the mechanism of optimisation is operated at saturation. 
\end{enumerate}   

We have analyzed the heat dissipation caused by kinetic proofreading in \Ecoli \; and we have argued that it constitutes a significant part of the heat measured in calorimetric experiments on this model organism. Further we have shown that the effect of the optimisation mechanism saturates and that Nature operates at saturation. 

In his 2025 Nobel Prize lecture Hopfield pointed out: ``\,`Free energy' is the key. High-energy molecules
like glucose or adenosine triphosphate are expensive for a cell to make, and a biochemical process using unexpectedly large numbers of high-energy molecules in a simple process must be paying that cost for a purpose.'' \cite{hopfield2025nobel} 

%  Suppose that purpose was to “proofread” to remove biosynthesis errors.

This argument can be taken further. Most of the energy contained in the currency metabolites is not transformed into chemical work, but dissipated as heat into an organism's environment. This might seem wasteful at first glance, but it provides living organisms with a flexible tool to optimise their fitness. 
That Nature has driven this mechanism into saturation means that living organisms do not make a compromise between a thrifty use of resources and a moderate improvement of chemical reaction networks. Rather, they take up all the energy needed to run the mechanism optimally even if most of the energy is `wasted' as heat. This indicates that from an evolutionary perspective, being able to adapt flexibly to various fitness functions is advantageous over being economical with energy consumption. If so, life is hot, because it is versatile and adaptive.

%TC:ignore
\section{Appendix}
\subsection{Values of $K/\mathcal{Q}$ for some key reactions in {\it E.~coli}} \label{app:sec:K/Q}

In Table~\ref{tab:K/Q}, we show $K/\cal{Q}$ for 24 reactions spanning the core of {\it E.~coli} central carbon metabolism, ordered from lowest to highest. The values have been calculated from measured Gibbs free energy data in Supplementary Table 4 of reference~\cite{Park2016}, and we have retained the acronyms used by these authors. The range given for each value, [low, high], reflects the error bars quoted by these authors. While some of these values are subject to large uncertainties, this table serves to explain what we mean by saying in the main text that the $K/\RQ\sim10^{9}$--$10^{12}$ for intracellular ATP hydrolysis is `very large'. 

% Requires: booktabs, siunitx
% In a two-column layout use table* (as here); \footnotesize keeps it within \textwidth.

\begin{table*}
\centering
\footnotesize
\caption{The $K/\mathcal{Q}$ ratio for 24 reactions associated with {\it E.~coli} central carbon metabolism (glycolysis, the pentose phosphate pathway, and the TCA cycle). Calculated from data in reference~\cite{Park2016}.}
\label{tab:K/Q}
\begin{tabular}{@{}llll@{}}
\toprule
Acronym & Pathway & Reaction & $K/\mathcal{Q}$ \\
\midrule
RPI     & PPP                & Ru5P $\rightarrow$ R5P                                   & $1.0016$ $[1.0012,\ 1.130]$                 \\
FUM     & TCA cycle          & Fum $\rightarrow$ Mal                                    & $1.0039$ $[1.0039,\ 1.0078]$                \\
RPE     & PPP                & Ru5P $\rightarrow$ Xu5P                                  & $1.0315$ $[1.0235,\ 1.0355]$                \\
TKT1    & PPP                & Xu5P + R5P $\rightarrow$ GAP + S7P                       & $1.168$ $[1.163,\ 1.177]$                   \\
MDH     & TCA cycle          & Mal + NAD $\rightarrow$ OAA + NADH                       & $1.223$ $[1.191,\ 1.277]$                   \\
TPI     & Glycolysis         & DHAP $\rightarrow$ GAP                                   & $1.358$ $[1.332,\ 1.385]$                   \\
GAPD    & Glycolysis         & GAP + NAD + P$_i$ $\rightarrow$ 13BPG + NADH            & $1.668$ $[1.000,\ 2.773]$                   \\
PGK     & Glycolysis         & 13BPG + ADP $\rightarrow$ 3PG + ATP                      & $1.734$ $[1.000,\ 2.773]$                   \\
PGI     & Glycolysis         & G6P $\rightarrow$ F6P                                    & $1.860$ $[1.681,\ 2.249]$                   \\
TKT2    & PPP                & Xu5P + E4P $\rightarrow$ GAP + F6P                       & $1.867$ $[1.748,\ 2.002]$                   \\
FBA     & Glycolysis         & FBP $\rightarrow$ DHAP + GAP                             & $2.155$ $[2.089,\ 2.240]$                   \\
ACONTa  & TCA cycle          & Cit $\rightarrow$ Icit                                   & $2.275$ $[1.115,\ 10.24]$                   \\
ENO     & Glycolysis         & 2PG $\rightarrow$ PEP                                    & $2.905$ $[1.000,\ 8.471]$                   \\
PGM     & Glycolysis         & 3PG $\rightarrow$ 2PG                                    & $3.419$ $[1.000,\ 8.471]$                   \\
TALA    & PPP                & S7P + GAP $\rightarrow$ E4P + F6P                        & $8.212$ $[7.541,\ 9.893]$                   \\
ICDHyr  & TCA cycle          & Icit + NADP $\rightarrow$ $\alpha$KG + NADPH + CO$_2$   & $10.01$ $[1.000,\ 398.4]$                   \\
PYK     & Glycolysis         & PEP + ADP $\rightarrow$ Pyr + ATP                        & $15.63$ $[7.198,\ 34.62]$                   \\
ME1     & Anaplerotic        & Mal + NAD $\rightarrow$ Pyr + NADH + CO$_2$              & $43.69$ $[22.77,\ 130.4]$                   \\
AKGDH   & TCA cycle          & $\alpha$KG + NAD + CoA $\rightarrow$ SuccCoA + NADH + CO$_2$ & $96.36$ $[7.779,\ \num{1.379e4}]$          \\
GND     & PPP                & 6PG + NADP $\rightarrow$ Ru5P + NADPH + CO$_2$           & $346.5$ $[30.46,\ \num{8.526e3}]$           \\
SUCOAS  & TCA cycle          & SuccCoA + ADP + P$_i$ $\rightarrow$ Succ + CoA + ATP     & $\num{4.177e3}$ $[55.78,\ \num{1.676e5}]$  \\
PFK     & Glycolysis         & F6P + ATP $\rightarrow$ FBP + ADP                        & $\num{1.450e4}$ $[\num{7.679e3},\ \num{2.666e4}]$ \\
PDH     & Pyruvate oxidation & Pyr + NAD + CoA $\rightarrow$ AcCoA + NADH + CO$_2$     & $\num{5.276e4}$ $[\num{2.371e3},\ \num{4.172e6}]$ \\
CS      & TCA cycle          & AcCoA + OAA $\rightarrow$ Cit + CoA                      & $\num{1.470e6}$ $[\num{1.700e4},\ \num{5.168e7}]$ \\
\bottomrule
\end{tabular}
\end{table*}

\subsection{Steady state solution}
\label{app:sec:HeatDiss}
To solve the chemical rate equations for the three state model depicted in fig.~\ref{fig:network}, we use the method described in ref.~\cite{schnakenberg1976network}. The steady state concentrations are
\begin{eqnarray}
\frac{c_1^\text{s}}{c_\text{tot}^\text{s}} & = & 
\frac{k_{12}k_{23} + k_{32}k_{13} + k_{12}k_{13}}{d}\\
\frac{c_2^\text{s}}{c_\text{tot}^\text{s}}& = & 
\frac{k_{21}k_{23} + k_{23}k_{31} + k_{21}k_{13}}{d}\\
\frac{c_3^\text{s}}{c_\text{tot}^\text{s}}& = & 
\frac{k_{21}k_{32} + k_{32}k_{13} + k_{12}k_{31}}{d}
\end{eqnarray}
with $c^\text{s}_{\text{tot}} := {c_1^\text{s} + c_3^\text{s} + c_3^\text{s}}$ and
\begin{equation*}
\begin{split}
&d := k_{21}k_{32} + k_{32}k_{13} + k_{13}k_{21} + k_{31}k_{23} + k_{23}k_{12}\\ 
&\hspace{3em}{} + k_{12}k_{31} + k_{13}k_{12}  +  k_{21}k_{23}  + k_{32}k_{31}\;.
\end{split}
\end{equation*}

The corresponding steady state current is
\[ J^\text{s} = \frac{c^\text{s}_\text{tot}}{d}\, (k_{21}k_{32}k_{13}- k_{12}k_{23}k_{31})\,.
\]

\subsection{Kinetic proof reading model}\label{app:sec:KPRModel}
To generate the results shown in Fig.~\ref{fig:kpr}, we implemented the following model, which hews closely to that described in Refs.~\cite{physicallens,physicallens2}.  The model comprises quasi-chemical reactions that represent: elongation factor (EF) complex formation (association), $\ce{C + GTP}\rightleftharpoons\ce{C.GTP}$\,; ribosomal binding $\ce{R + C.GTP}\rightleftharpoons\ce{R.C.GTP}$\,; hydrolysis, $\ce{R.C.GTP}\rightleftharpoons\ce{R.C.GDP + Pi}$\,; detachment, $\ce{R.C.GDP}\rightleftharpoons\ce{R + C.GDP}$\,; and dissociation, $\ce{C.GDP}\rightleftharpoons\ce{C + GDP}$.  All reactions are assumed reversible, though out of equilibrium some are strongly driven in the forward direction.
Rate coefficients are, respectively, in the forward and back directions, $g_t'=\kon\equiv10^8\,\mathrm{M}^{-1}\,\mathrm{s}^{-1}$, $g_t=\koff\equiv100\,\mathrm{s}^{-1}$\,; $\kC'=\kon$, $\kC=\koff$\,; $m'=0.1\,\koff$, for $m$, see below; $\lC=\koff$, $\lC'=0.01\,\kon$\,; $g_d=10\,\koff$, $g_d'=\kon$.  A so-called `cycle constraint' fixes the value of $m$ in the hydrolysis step according to $m'/m = \KGTP (\lC'/\lC)(\kC/\kC')(g_t/g_t')(g_d'/g_d)$, where $\KGTP$ is the equilibrium constant for the hydrolysis equilibrium $\ce{GTP}\rightleftharpoons\ce{GDP + Pi}$\,; we take $\KGTP=5\times10^5\,\mathrm{M}$ so that $m=0.02\,\mathrm{M}^{-1}\,\mathrm{s}^{-1}$.  The cycle constraint ensures that the cycle flux vanishes when the concentrations of GTP, GDP and inorganic phosphate (Pi) are at their equilibrium values.  We do not explicitly represent polypeptide elongation in the model since it has a very small rate compared to the cycle reactions; rather this step is assumed proportional to the steady-state concentration of the hydrolysed bound EF complex $[\ce{R.C.GDP}]$.  We also omit direct hydrolysis of \ce{GTP} and \ce{C.GTP} since the rates can be assumed negligible under the relevant conditions. Finally, as in Ref.~\cite{physicallens}, a parallel set of reactions with `D' representing the attachment of a `wrong' amino acid are also included, with detachment rate coefficients $\kD$ and $\lD$ increased by a factor $f=100$ compared to `C'\,; the translation error rate is then defined as the ratio $[\ce{R.D.GDP}]/[\ce{R.C.GDP}]$.
We assume the currency metabolites are chemiostatted as $[\ce{GTP}]=[\ce{G}][\ce{Pi}]/([\ce{Pi}]+\QGTP)$ and $[\ce{GDP}]=[\ce{G}]\QGTP/([\ce{Pi}]+\QGTP)$, where $\QGTP=[\ce{GDP}][\ce{Pi}]/[\ce{GTP}]$ is the reaction quotient for the notional GTP hydrolysis equilibrium and $[\ce{G}]=[\ce{GTP}]+[\ce{GDP}]$ is the held-constant phosphorylation-state agnostic total concentration of the currency metabolite.  To obtain the results shown in Fig.~\ref{fig:kpr}, we set the total individual concentrations of C and D at $0.1\,\mathrm{mM}$, calculate the total ribosomal concentration assuming a copy number of $5000$ ribosomes in a volume $\SI{2}{\micro\meter^3}$, take $[\ce{G}]=[\ce{Pi}]=1\,\mathrm{mM}$ as typical physiological values, and solve for the non-equilibrium steady state (NESS) as a function of the degree of disequilibrium by varying the reaction quotient $\QGTP$ between $5\times10^{-7}\,\mathrm{M}$ and the equilibrium  upper bound $\KGTP=5\times10^5\,\mathrm{M}$.  Different choices here all result in similar phenomenology.  We use the BASICO interface~\cite{Bergmann2023} to the COPASI simulation platform~\cite{Hoops2006} to perform the actual calculations.  A Python script which implements the model and generates Fig.~\ref{fig:kpr} is available on request.

\subsection{Flux-balance analysis}\label{app:sec:FBAAnalysis}
We used two recent genome-scale metabolic reconstructions, iJO1366~\cite{Orth2011} and iML1515~\cite{Monk2017}, which represent comprehensive models of the metabolic capabilities of \Ecoli, to examine the ATP maintenance demands.  We adopted the default growth medium which comprises glucose as the sole carbon/energy source; ammonium, inorganic phosphate and sulfate ions to satisfy the elemental demands for N, P, and S in the biomass; and sundry trace nutrients and minerals unimportant for present purposes.  Under aerobic conditions, and limiting the glucose uptake rate to $10\,\mathrm{mmol}\,\mathrm{gDW}^{-1}\,\mathrm{hr}^{-1}$, we optimise for the flux to biomass using flux-balance analysis (FBA)~\cite{Palsson2010, palsson2015}.  This yields growth rates of 0.98--$0.88\,\mathrm{hr}^{-1}$, where here and in the main text the figure for iJO1366 is given first.  We then interrogate the models to determine the grow-associated ATP maintenance (GAM) demand incorporated in the flux to biomass, and the non-growth associated ATP maintenance (NGAM) demand associated to the reaction $\ce{ATP}+\ce{H2O}\rightarrow\ce{ADP}+\ce{H+}+\ce{Pi}$.  Combining these gives the result quoted in the main text.  
To do the heat production calculation we assume a reaction enthalpy $\Delta H\simeq -30.5\,\mathrm{kJ}\,\mathrm{mol}^{-1}$ for ATP hydrolysis, take the dry mass of an \Ecoli\ bacterium as $\simeq0.43\,\mathrm{pg}$, and divide by the doubling time to get the quoted heat production rates.  
The models were downloaded from the BiGG Models database~\cite{King2016}, and used as given.
FBA calculations were performed using the COBRApy platform~\cite{Ebrahim2013}.  
The Python scripts which perform the calculations are available on request.

%TC:endignore
%TC:ignore
\begin{acknowledgments}
WCKP thanks the Isaac Newton Institute for Mathematical Sciences, Cambridge for support and hospitality during the 2023 programme ‘New statistical physics in living matter’ (supported by EPSRC grant EP/Z000580/1), where the questions that led to the current work were first conceived. TS thanks the Higgs Centre for Theoretical Physics, The University of Edinburgh, for part funding a sabbatical visit to Edinburgh, where the research was initiated and substantially completed. We thank Anja Seegebrecht, Thorsten Friedrich, Thorsten Hugel, Daan Frenkel, Andreas H\"artel and Stefan Miller for feedback on the manuscript.
\end{acknowledgments}
%TC:endignore

\bibliography{LifeIsHot}% Produces the bibliography via BibTeX.

\end{document}